\documentclass[12pt]{article}

\usepackage{fancyhdr}
\usepackage{epsfig}
\usepackage{cite}
\usepackage{graphicx}
\usepackage[centertags]{amsmath}
\usepackage{theorem}

\usepackage{graphicx}
\usepackage{amssymb}
\usepackage{latexsym}
\usepackage{pstricks}
\usepackage{color}
\usepackage{mathrsfs}
\usepackage{latexsym}
\usepackage{braket}
\usepackage{slashed}
\usepackage{bbold}
\newcommand{\be}{\begin{equation}}
\newcommand{\ee}{\end{equation}}
\newcommand{\bea}{\begin{eqnarray}}

\newcommand{\eea}{\end{eqnarray}}

\numberwithin{equation}{section}
\def\hybrid{\topmargin 22pt    \oddsidemargin 0pt 
      \headheight 0pt \headsep 0pt
      \textwidth 6.5in        
       \textheight 9in         
      \marginparwidth .875in
      \parskip 5pt plus 1pt   \jot = 1.5ex}

\hybrid

\usepackage{caption} 
\newcommand{\refe}[1]{Eqn.~(\ref{#1})}

\renewcommand{\thefootnote}{\fnsymbol{footnote}}

\begin{document}
\center{
\begin{flushright} 
DESY 13-048\\
DOI:10.1007/JHEP04(2013)161
\end{flushright}
\vspace{1cm}
\begin{center}
{\Large\textbf{5D Maximally Supersymmetric Yang-Mills\\ in 4D Superspace: Applications}}
\vspace{1cm}

\textbf{Moritz McGarrie$^{1,}$\footnote[2]{\texttt{moritz.mcgarrie@desy.de}  }
}\\
\end{center}

{
\it{ ${}^1$ 
Deutsches Elektronen-Synchrotron,\\ DESY, Notkestrasse 85, 22607 Hamburg, Germany
}\\

}
\setcounter{footnote}{0}
\renewcommand{\thefootnote}{\arabic{footnote}}

\abstract{We reformulate 5D maximally supersymmetric Yang-Mills in 4D Superspace, for a manifold with boundaries. We emphasise certain features and conventions necessary to allow for supersymmetric model building applications.  Finally we apply the holographic interpretation of a slice of AdS and show how to generate Dirac soft masses between external source fields, as well as kinetic mixing, as a boundary effective action. }

\section{Introduction}
Convincing evidence has been shown that maximal super Yang-Mills in 5D is a good  effective description of the 6D $(2,0)$ M5-brane CFT on an $S^1$, for example see \cite{Douglas:2010iu,Lambert:2010iw}.  This suggestion is interesting from the perspective of beyond standard model building.  Whilst we usually think of quantum field theories as effective field theories, with some perhaps unmentioned cutoff, extra dimensional models suffer more than most from their lack of UV completion as they appear, at least from power counting,  to be non renormalisable.  If 5D MSYM is indeed UV finite, after accounting  for both the perturbative and non-perturbative spectrum, then 
it may be possible to build 5D models which have a self contained UV completion.

Precedents for this sort of model building are well known:  The Horava-Witten construction in eleven dimensional supergravity \cite{Horava:1996ma} motivated studies of 5D global super Yang-Mills with boundaries \cite{Mirabelli:1997aj}. Later Randall-Sundrum \cite{Randall:1999ee} models motivated warped or ``a slice of'' $AdS_5$ models.  AdS/QCD constructions \cite{Erlich:2005qh,DaRold:2005ju} have also flourished as they capture the relevant local quantum field theory degrees of freedom of string theory, in five dimensions. Already in \cite{Douglas:2010iu,Lambert:2010iw} a $T^2$ compactification of the $(2,0)$ theory was used to consider four dimensional physics.  As the Torus is the product of two circles  $T^2= S^1\times S^1$ with Radii $R_5$ and $R_4$, naturally the 5D MSYM action is the limit that $R_5$ is small relative to $R_4$.    We would like to explore the resulting action where $R_4$ is an orbifold $S^1/\mathbb{Z}_2$.  In 4d superspace notation this leads to a vector superfield and chiral adjoint with positive parity ($V,H$) and two chiral superfields with negative parity ($\Phi,H^c$).  In some sense this model is already in use for model building:  whenever one wishes for a positive parity vector $V$ and $H$ for instance to generate Dirac soft masses between the fermions contained in those multiplets as in \cite{Abel:2011dc,Benakli:2012cy},  it may be natural to consider MSYM in 5D and not simply SYM in 5D.  Furthermore in the examples we choose despite $\Phi$ and $H^c$ having negative parity, they actually play a signification role, due to the equations of motion between the left and right handed fermions of each multiplet.  Additional chiral adjoints  such as ($H_{adj}$) of $SU(3)_c$ are indeed well motivated extensions which have been shown to be able to increase the Higgs mass substantially \cite{Bhattacharyya:2012qj}. Indeed theories with more supersymmetry may fair better than models with less,  as discussed in \cite{Heikinheimo:2011fk}. Brane constructions, such as those found in this paper, offer the possibility that matter multiplets appear in $\mathcal{N}=1$ multiplets and gauge fields in $\mathcal{N}=2$ multiplets, something that arises quite naturally in supersoft supersymmetry breaking.

In this paper we do not concern ourselves with the question of UV finiteness, instead wish to study this setup from the perspective of model building. A natural first step to make use of the 5D MSYM action for model building purposes is to reduce this action to 4D superspace, in two component spinor notation.  Conventions are important here: we have chosen the gamma matrices of the various dimensions to be built from natural tensor products of the four dimensional Weyl representation.  In this paper we will compactify on an interval $S^1/\mathbb{Z}_2$ the $x_4$ direction, preserving $x_5$ as the direction of M-theory, and hence hopefully preserving some of the UV-finiteness of this theory.  In particular the choice of an interval instead of a circle allows for the possibility to break some supersymmetry, leaving a theory with only  8 supercharges from the four dimensional perspective.  In addition the orbifold fixed points allows one to introduce boundary localised matter, which may break the  supersymmetry down further, hopefully in which the final  supersymmetry (last 4 real supercharges) can be broken dynamically.

The current paper establishes the procedure for reducing MSYM in 5D starting from 11D spinors, to four dimensional $\mathcal{N}=1$  superspace notation, with two component spinors.   The reduction of  $\mathcal{N}=1$, 5D SYM has been carried out in \cite{Mirabelli:1997aj,Hebecker:2001ke,ArkaniHamed:2001tb}.  The procedure, starting from $\mathcal{N}=1$, 7d SYM,
may be found in \cite{Ludeling:2011ip}.  Recent papers on supersymmetry breaking and model building using the 5D SYM action may be found in \cite{McGarrie:2010yk,McGarrie:2010kh,Abel:2010vb,McGarrie:2011av,McGarrie:2012fi}.  It is also interesting to consider their deconstruction \cite{ArkaniHamed:2001ie,McGarrie:2010qr,Abel:2011dc,McGarrie:2011dc,Lambert:2012qy}.

This structure of the paper is as follows:
In section \ref{11daction} we outline the maximal super Yang-Mills action in five dimensions written using eleven dimensional spinors.  In section \ref{symplecticmaj}  we construct the same action using symplectic Majorana spinors, which are the natural spinors of five dimensions. In section  \ref{4dsuperspace} we start from section \ref{symplecticmaj}  and take an orbifold of the fifth direction $x_4$, constructing the resulting lagrangian and notation for the theory in superspace. In section \ref{Dirac} we demonstrate an application of this setup by generating an effective action for Dirac soft masses between boundary external sources. We conclude in section \ref{Conclude}. A very convenient set of conventions may be found in appendix \ref{conventions}, which also briefly comments on this construction's descent from $\mathcal{N}=1$ super Yang-Mills in seven dimensions.

\section{5D maximal super Yang-Mills} \label{11daction}
In this section we outline the maximal super Yang-Mills action in five dimensions written using eleven dimensional spinors living in the spacetime of M-theory.   

The 5D maximal super Yang-Mills action \cite{Lambert:2010iw} is given by
\be
S_{MSYM}=\int d^5 x \mathcal{L}_{MSYM}
\ee
where the lagrangian  $\mathcal{L}_{MSYM}$ is given by
\be
\text{tr}\! \left(\! -\frac{1}{4}F_{MN}F^{MN}\!-\! \frac{1}{2}D_{M}X^{I}D^{M}X^{I}+\frac{i}{2}\bar{\Psi}\Gamma^{M}D_{M}\Psi+\frac{ig_5}{2}\bar{\Psi}\Gamma^5 \Gamma^I [X^I,\Psi] +\frac{g^2_5}{4}\sum_{I,J}[X^I,X^J]^2 \right)
\ee
The indices $\mu,\nu$ are 4D, $\mu=0,1,2,3$.  $M,N$ are 5D indices with metric $\eta_{MN}=\text{diag}(-1,1,1,1,1)$.  $A,B$ are the 11D indices.  $I,J=6,7,8,9,10$.  The spinors are real with 32 components. 
 $\bar{\Psi}=\Psi^{\dagger}\Gamma^0$.  We have chosen to define $D_M= \partial_M+ig_5 A_M$, which rescales the $1/g^2_{5}$ inside the field strength tensor.  In this non canonical normalisation\footnote{This form of the action is more suited to perturbation theory in $g_5$, the canonical form being most suited to finding solitonic solutions.},  the mass dimensions of $[D_M,A_M,X^I,\Psi]$ are $[1,3/2,3/2,2]$. Similarly $[\delta_{\epsilon},\epsilon, \theta ,Q]$ has $[0,-\frac{1}{2},-\frac{1}{2},\frac{1}{2}]$.  The complete conventions may be found in appendix \ref{conventions}.

The supersymmetry transformations are given by
\be
\delta_{\epsilon}A_{M}=i\bar{\epsilon}\Gamma_{M}\Gamma_5\Psi
\ee
\be
\delta_{\epsilon}X^I=i\bar{\epsilon}\Gamma^I \Psi
\ee
\be
\delta_{\epsilon}\Psi=\frac{1}{2}F_{MN}\Gamma^{MN}\Gamma_5 \epsilon+D_M X^I \Gamma^M \Gamma^I \epsilon-\frac{ig_5}{2}[X^I,X^J]\Gamma^{IJ}\Gamma^5\epsilon
\ee
where $\epsilon$ is a supersymmetry transformation parameter. Useful identities are 
\be
\Gamma^{MN}=\frac{1}{2}[\Gamma^M,\Gamma^N]
\ee
\be
D_{M}X^I= \partial_{M}X^I-i g_5[A_M,X^I]
\ee
\be
F_{MN}=\partial_MA_N-\partial_N A_M-ig_5[A_N,A_M].
\ee
This theory has 16 supercharges,
\be
\Psi=\Psi_+ +\Psi_-
\ee
\be
\Psi_{\pm}=\pm\Gamma^5 \Psi_{\pm}
\ee
\be
\epsilon_{\pm}=\mp\Gamma^5 \epsilon_{\pm}.
\ee
For more details see \cite{Ludeling:2011ip}.   The action is written in terms of full 11D M-theory spinors, which are unconstrained real 32-component spinors, despite only the projection of the spinor with respect to $\Gamma_5$ actually living on the M5-brane.

\subsection{Important boundary terms}
We will wish to study the inclusion of supersymmetric matter fields in both the orbifold and boundary perspectives.   In the orbifold picture one can neglect total derivatives in satisfying the supersymmetry transformations.  The are however two types of boundary conditions that must be satisfied for delta-function localised matter: primary boundary conditions make the bulk and boundary supersymmetric and secondary boundary conditions also make the boundary conditions  themselves supersymmetric.  These are naturally included in the superspace formulation, but must be introduced to the component action to complete supersymmetry.  These terms are of  particular interest as, in the boundary picture, they are the boundary terms for this action (analogous to the Gibbons-Hawking-York boundary terms of gravity  as discussed in  \cite{Belyaev:2005rs,Belyaev:2006jg}). For an example of completing actions with the correct supersymmetric boundary term see \cite{Berman:2009kj} and also for double field theory \cite{Berman:2011kg}.

These terms play a significant role in AdS systems, where this action determines the boundary to boundary two point functions using the operator field correspondence, for a given bulk action with boundary sources \cite{Arutyunov:1998ve,McGarrie:2012fi}.

On a manifold with a boundary at $x_4=0$ one must in addition include the boundary terms (in canonical normalisation)
\be
\frac{1}{g_5^2}\int_{\partial \mathcal{M}} d^4 x \left( \eta^{4M} \eta^{PQ}F_{MP}A_{Q}-\eta^{4M}(D_M X_I)X^I +\frac{1}{2}\bar{\Psi}\Psi \right)
\ee
for full closure of supersymmetry of the bulk action.  Strictly speaking it is actually the second two terms that are the additional boundary terms, the first term being already contained in the Yang-Mills action, but we include them together because of their equivalent roles in the boundary action.  As we shall see later, the superfield notation automatically includes these boundary terms and are related to the boundary conditions of the bulk fields.

\section{As symplectic Majorana spinors}\label{symplecticmaj}
In this section we define the maximal super Yang-Mills action using symplectic Majorana spinors, which are the natural objects of five dimensions.  This action has and $SO(5)_R$ symmetry and no additional auxiliary fields. This action may be compared with that of $\mathcal{N}=1$ SYM \cite{Hebecker:2001ke}, in which the action has an $SU(2)_R$ symmetry and a triplet of auxiliary fields $X^a$, under the $SU(2)_R$.

The lagrangian is given by 
\begin{equation}
\mathcal{L}_{MSYM}=
\begin{aligned}[t]
&-\frac{1}{4}F_{MN}F^{MN}-\frac{1}{2}D_{M}X_{I}D^{M}X^{I}+\frac{i}{2}\bar{\psi}^I\Gamma^{M}D_{M}\psi_I    \\
& +\frac{g^2_5}{4}\sum_{a,b}[X^a,X^b]^2 + \frac{g_5}{2}\bar\psi^I(G^a)_I\phantom{}^J[X_a,\psi_J]
\end{aligned}
\end{equation}
where the trace, $\text{tr}$, is implicit.  The $I,J$ label the full $SP(2)$ R-symmetry, which decomposes into two copies of the usual $SU(2)$ R-symmetry for 5D symplectic Majorana spinors, labelled by $i,j$.  Incidentally, $\bar{\epsilon}_I=(\epsilon_I)^{\dagger}\Gamma^0$.

We define a basis of fermions
\begin{equation}
\label{eq:fermion}
 \psi_I=\left(
 \begin{array}{c}
   \Psi^1\\
   \Psi^2\\
   \Omega^1\\
   \Omega^2
 \end{array}\right),\qquad
\left(\bar{\psi}_I\right)^T=\left(
 \begin{array}{c}
  \bar{\Psi}_1\\
    \bar{\Psi}_2\\
     \bar{\Omega}_1\\
\bar{\Omega}_2
 \end{array}\right),\qquad
 \epsilon_I=\left(
 \begin{array}{c}
   \epsilon^1\\
   \epsilon^2\\
   \xi^1\\
   \xi^2
 \end{array}\right),
\qquad
\left( \bar{\epsilon}_I\right)^T=\left(
 \begin{array}{c}
  \bar{ \epsilon}_1\\
  \bar{ \epsilon}
_2\\
  \bar{ \xi}_1\\
  \bar{ \xi}_2
 \end{array}\right).
\end{equation}
The supersymmetry transformations are given by
\begin{subequations}
\label{eq:susytransfo}
\begin{align}
  \delta_\epsilon X^a & = i{\bar\epsilon}^I(G^a)_I\phantom{}^J\psi_J\\
  \delta_{\epsilon}A_{M}& =i\bar{\epsilon}_{i}  \gamma_{M}\Psi^i_{}+ i\bar{\xi}_{i}  \gamma_{M}\Omega^i\\
  \delta \psi_I & = F_{MN}\Gamma^{MN}\epsilon_I+ \slashed{D}X^a \left(G^a \right)_I\phantom{}^J\epsilon_J+\frac{g_5}{2}[X_a,X_b](G^{ab})_I\phantom{}^J\epsilon_J
\end{align}  
\end{subequations}
The super algebra may be written as 
\be
\{Q^i_{A},\bar{Q}^j_{B}\}=2\Gamma^M P_M \delta^{ij}\delta_{AB}
\ee
$A,B=1,2$ where the Q's labelled $A=1$ couple to $\epsilon^i$ and those labelled $A=2$ to $\xi^i$.
We label
\begin{equation}
X^a=X^6,\dotsc,X^{10}
\end{equation}
while the matrices $G^a$ are explained below in section \ref{sec:so5-r-symmetry}.
We will not need the full R-symmetry in the following discussions and therefore we singled out half of the fermions in writing \eqref{eq:fermion}.

We also briefly comment that we could have written the action using unconstrained Dirac spinors (of five dimensions). The resulting action would contain
\be
\mathcal{L}\supset \frac{i}{2}\bar{\lambda}\Gamma^MD_M\lambda+ \frac{i}{2}\bar{\chi}\Gamma^MD_M\chi
\ee
where $\lambda=\Psi^1$ and $\chi=\Omega^1$ as in \refe{spinors}. Although in this case the R-symmetry would not have been so manifest. 

\subsection{The boundary term}
The 5D maximal super Yang-Mills boundary term with symplectic Majorana spinors is 
\be
\frac{1}{g_5^2}\int_{\partial \mathcal{M}}d^4 x \left( G^{4M} G^{PQ}F_{MP}A_{Q}-\frac{1}{2}G^{4M}(D_M X_a)X^a +\frac{1}{4}\bar{\psi}_I\psi^I \right).
\ee
It is straightforward to determine from this the boundary terms in 2 component spinor notation.

\subsection{The $SO(5)$ R-symmetry}
\label{sec:so5-r-symmetry}
The 5D maximally supersymmetric Yang-Mills theory has $SO(5)\cong Sp(2)$ R-symmetry. The two pairs of symplectic-Majorana fermions transform as $\mathbf{4}$ of $Sp(2)$ whereas the five scalars transform as $\mathbf{5}$ of $SO(5)$. These two representations can be related by using five-dimensional, Euclidean gamma matrices
\begin{equation}
  \left(G^a\right)_I\phantom{}^J\! \! = \! \left\{
    \left(\!\!
      \begin{array}{cc}
        0&-i\sigma^3\\
        i\sigma^3&0
      \end{array}\!\!
    \right),
    \left(\!\!\!
      \begin{array}{cc}
        0&\mathbb{1}_2\\
        \mathbb{1}_2&0
      \end{array}\!\!\!
    \right) ,
    \left(\!\!
      \begin{array}{cc}
        0&-i\sigma^2\\
        i\sigma^2&0
      \end{array}\!\!
    \right) ,
    \left(\!\!
      \begin{array}{cc}
        0&-i\sigma^1\\
        +i\sigma^1&0
      \end{array}\!\!
    \right),
    \left(\!\!
      \begin{array}{cc}
        \mathbb{1}_2&0\\
        0&-\mathbb{1}_2
      \end{array}\!\!
    \right)
\right\}.
\end{equation}
They satisfy
\begin{equation}
  \{G^a,\bar G^b\} = -2 \delta^{ab}, \qquad a,b=6,\dotsc,10.
\end{equation}
The index $a$ relates to the scalar components  $X^a$  for which $X^{10}\equiv \Sigma$.

\section{In terms of 4D superspace}\label{4dsuperspace}
We now wish to rewrite the maximal super Yang-Mills description in terms of four dimensional superfields.   For dimensional reduction to four dimensional superspace, it is more natural to first formulate a description using 5D symplectic Majorana spinors, which each decompose into two 4D Weyl spinors. This will also make the R-symmetry more manifest.  The spinors of \refe{eq:fermion} are given by 
\be
\Psi^1 = \left(\begin{array}{c}
\lambda_{L \alpha}\\ 
\bar\lambda_{R}^{\dot\alpha}
\end{array}\right)~,~~~
\Omega^1 = \left(\begin{array}{c}
\chi_{L \alpha}\\ 
\bar{\chi}_{R }^{\dot{\alpha}}
\end{array}\right)~,~~~
\Psi^2 = \left(\begin{array}{c}
\lambda_{R\alpha}\\ 
-\bar\lambda_{L}^{\dot\alpha}
\end{array}\right)~,~~~
\Omega^2   = \left(\begin{array}{c}
\chi_{R\alpha}\\
-\bar{\chi}_{L}^{ \dot{\alpha}}
\end{array}\right)~.\label{spinors}
\ee
The reality condition defines the barred fermions by
\be
\Psi^i=\epsilon^{ij}C_5\bar{\Psi}_{j}^T \ \ \ \text{and} \ \ \ \Psi^i=\epsilon^{ij}B_5\Psi^*_{j}.
\ee
where the $SU(2)_R$ symmetry indices are raised and lowered with 
\be
 \epsilon^{ij}= \left(\!\!
      \begin{array}{cc}
        0&1\\
        -1&0
      \end{array}\!\!
    \right).
\ee
The supersymmetry transformation parameters $\epsilon^i, \xi^i$ are defined similarly:
\be
\epsilon^1 = \left(\begin{array}{c}
\epsilon_{L \alpha}\\ 
\bar\epsilon_{R}^{\dot\alpha}
\end{array}\right)~,~~~
\xi^1 = \left(\begin{array}{c}
\xi_{L \alpha}\\ 
\bar{\xi}_{R }^{\dot{\alpha}}
\end{array}\right)~,~~~
\epsilon^2 = \left(\begin{array}{c}
\epsilon_{R\alpha}\\ 
-\bar\epsilon_{L}^{\dot\alpha}
\end{array}\right)~,~~~
\xi^2   = \left(\begin{array}{c}
\xi_{R\alpha}\\
-\bar{\xi}_{L}^{\alpha}
\end{array}\right)~.
\ee
For these spinors $\bar{\epsilon}_1=(\epsilon^1)^{\dagger}\gamma^0= (\epsilon^{\alpha}_R,\bar{\epsilon}_{L,\dot{\alpha}})$ and $\bar{\epsilon}_2= (-\epsilon^{\alpha}_L,\bar{\epsilon}_{R,\dot{\alpha}})$.  It will be useful to label $\Sigma\equiv X^{10}$.  

We are actually interested to explore manifolds with boundaries, as they have the most useful practical applications. The presence of constant boundaries preserves only half the supersymmetry of the bulk system.  As the commutator of a supersymmetry transformation generates a translation, we may define the translation parameter $a^M$ in terms of the supersymmetry transformation parameters,
\be
a^M=2i(\bar{\epsilon}_I \Gamma^M \eta^I).
\ee
Allowing only $a^5=0$ to break translation invariance, fixes a relation between the supersymmetry transformations that
\be
2i (\bar{\epsilon}_1\gamma^5\eta^1-\bar{\epsilon}_2\gamma^5\eta^2-\bar{\xi}_1\gamma^5\tilde{\eta}^1+\bar{\xi}_2\gamma^5\tilde{\eta}^2)=0
\ee
\be
(\epsilon_R\eta_L-\bar{\epsilon}_L\bar{\eta}_R-\epsilon_L\eta_R+\bar{\epsilon}_R\bar{\eta}_L)
-(\xi_R\tilde{\eta}_L-\bar{\xi}_L\bar{\tilde{\eta}}_R-\xi_L\tilde{\eta}_R+\bar{\xi}_R\bar{\tilde{\eta}}_L)=0
\ee
This means we can either make the $\epsilon$'s be related to the $\xi$'s i.e. $\epsilon^i=\beta \xi^i$, or we can make $\epsilon_L =\beta \epsilon_R$ and $\xi_L =\beta \xi_R$.  In the first case the $\epsilon$'s would be preserved and we could set $\xi_i=0$.   To solve the coupled fermion equations of motion, it turns out to be more practical to use $\epsilon_R=\xi_R=0$.  This second case is actually more familiar as it allows for parity ($+,-$) to be related to handedness ($L,R$).

Setting $\epsilon_R=\xi_R=0$, preserves only $\epsilon_{L}$ and $\xi_{L}$ or 8 real supercharges of $\mathcal{N}=2$ supersymmetry.  We may also temporarily set $\xi_L=0$ to obtain $\mathcal{N}=1$ multiplets. The positive parity fields fill a vector multiplet $V$ and a chiral multiplet $H$:  the field content of the preserved $\mathcal{N}=2$ SYM.  The negative parity fields fill two chiral multiplets $\Phi$ and $H^c$ in the adjoint which amount to an $\mathcal{N}=2$ Hypermultiplet.  This matter content also has a natural descent from applying a quiver to 4D  $\mathcal{N}=4$ super Yang-Mills \cite{Kirsch:2003kx,Kirsch:2004km}.
\subsection{The Lagrangian}
In order to write the action for the 5D maximally supersymmetric Yang-Mills theory in the 4D superspace, we need to collect the field content in $\mathcal{N}=1$ multiplets. In other words, we need to show that, after specialising to a $\mathcal{N}=1$ subset of the full supersymmetry, the fields, or certain linear combinations of them, transform as components of three chiral superfields and a vector superfield.

The terms in the Lagrangian for the Vector superfield and Chiral field $\Phi$ is
\be
\mathcal{L}= \frac{1}{2 }\text{tr}\int d^2 \theta W^{\alpha} W_{\alpha}  +\int d^2\bar{\theta} \bar{W}_{\dot{\alpha}}\bar{ W}^{\dot{\alpha}}+ \frac{1}{2g^2_{5}}\int d^4 \theta \  \text{tr}Z^2
\ee
with 
\begin{equation}
  Z = e^{-2g_{5}V}\left(\partial_4e^{2g_{5}V}+ig_{5}\bar{\Phi}e^{2g_{5}V}-ig_{5}e^{2g_{5}V}\Phi \right).
\end{equation}
The field strength tensor $W^{\alpha}$ is a left handed chiral Superfield defined by $W^{\alpha}=-\frac{1}{4}\bar{D}^2D^{\alpha}V$.
The lagrangian for the additional adjoint fields are given by
\be
 \mathcal{L}=\int d^4\theta \text{tr}\left( \  e^{-g_5 V} H^\dagger e^{g_5 V}H+  \  e^{-g_5 V} ( H^c)^{\dagger} e^{g_5 V}H^{c}  \right)\nonumber
\ee
\be
\ \ \ \ \ \ \ \ \ \ \ \ +\frac{1}{4} \int d^2\theta  \ \text{tr}\left( H^c \partial_4 H +g_5\Phi[H,H^c] \right)+c.c.
\ee
The gauge transformations are given by 
\be
e^{-V}\rightarrow e^{-\Lambda}e^{-V} e^{-\Lambda^{\dagger}}  \ \ \ ,  \ \ \  H^{\dagger}\rightarrow e^{\Lambda^{\dagger}}H^{\dagger}e^{-\Lambda^{\dagger}}.
\ee
The vector superfield in Wess-Zumino gauge is
\be
V=-\theta \sigma^{\mu}\bar\theta A_{\mu} +i\theta^2\bar{\theta}\bar{\lambda}_{L}-i\bar{\theta}^2\theta\lambda_{L}+\frac{1}{2}\theta^2\bar{\theta}^2( D )
\ee
where $D= D_4\Sigma=D_4 X^{10}$. The adjoint chiral Superfield is 
\be
\Phi=(\Sigma+i A_4)+\sqrt{2}\theta(-i\sqrt{2}\lambda_{R})+\theta^2 (F_{\Phi})
\ee
These field assignments within the multiplet are determined by the preserved supersymmetry transformations, below.
We wish to choose the boundary conditions following table \ref{table1},
\begin{table}
\begin{center}
\begin{tabular}{|c|c|c|c|c|}
\hline
Field & $V$ & $\Phi$ & $H$ & $H^c$\\
\hline
Parity & $+$ & $-$&$ +$& $-$ \\
\hline
\end{tabular}
\caption{The parity assignments of the bulk fields}\label{table1}
\end{center}
\end{table}
and so have chosen to preserve $\epsilon_L$ and $\xi_L$ and set $\epsilon_R=\xi_R=0$ and then $P(\partial_5)=-1$ as normal.  The resulting 4D $\mathcal{N}=2$ vector multiplet is the combination $V+H$ instead of $V+\Phi$.

It is  also instructive to see that if we choose to preserve only $\epsilon_L$ and set $\epsilon_R=0$ we may then break the symmetry down to $\mathcal{N}=1$ SYM and three adjoint chiral superfields.
In the orbifold direction $x_4$ there is still a residual gauge transformation and we can construct a super gauge covariant derivative operator in this direction
\be
\nabla_4\equiv \partial_4+  g_5\Phi \ \  \text{where} \ \ \nabla_4(\cdot) \equiv \partial_4(\cdot)+ g_5 \Phi (\cdot)-g_5 (\cdot)\Phi,
\ee
when acting on chiral objects and 
\be
\nabla_4(\cdot) \equiv \partial_4(\cdot) - g_5 \Phi^{\dagger} (\cdot)-g_5 (\cdot)\Phi,
\ee
when acting on real linear objects e.g. $(\cdot)=e^{2V}$.
The chiral super field strength tensor is
\be
W_{\alpha}=-i\lambda_{L}+ \sqrt{2}\theta_{\beta} (\delta^{\beta}_{\alpha} D- (\sigma^{\mu\nu})^{\beta}_{\alpha}F_{\mu\nu})+\theta^2 \sigma^{\mu}\partial_{\mu}\bar{\lambda}_{L}
\ee
 and the additional adjoint fields are 
\be
  H=(X^8+ iX^{9})+\sqrt{2}\theta (-i\sqrt{2}\chi_L)+\theta^2 (F_{H})
\ee
\be
  H^c=(X^6+iX^7)+\sqrt{2}\theta (\sqrt{2}\chi_R)+\theta^2 (F_{H^c}).
\ee
The F-terms of the chiral fields are given by
\be
F^{\dagger}_{\Phi}=-\frac{1}{2} g_5[(X^6+iX^7),(X^8+iX^9)]
\ee
\be
F^{\dagger}_{H}= -\frac{1}{2}\left[\partial_4(X^6+iX^7) +g_5[(\Sigma+iA_4)  , (X^6+iX^7)]  \right]
\ee
\be
F^{\dagger}_{H^{c}}=- \frac{1}{2}\left[ \partial_4(X^8+iX^9) +  g_5[(\Sigma+iA_4) ,  (X^8+iX^9)]   \right].
\ee
If one chooses $\xi_L=\xi_R=\bar\xi_L=\bar\xi_R=\epsilon_R=\bar\epsilon_R=0$ and considers only the supersymmetry transformations parameterised by $\epsilon_L$ and $\bar\epsilon_L$, in other words only $\mathcal{N}=1$ in four dimensions,  the general transformations \eqref{eq:susytransfo}   for the scalars reduces to
\begin{subequations}
  \begin{align}
\mathcal{N}&=2 & \mathcal{N}&=1 \nonumber \\
\delta X^6&= \chi_R \epsilon_L + \bar\chi_R\bar\epsilon_L-\bar{\lambda}_R\bar{\xi}_L -\lambda_R\xi_L&  \delta_{\epsilon_L} X^6&=  \chi_R \epsilon_L + \bar\chi_R\bar\epsilon_L\\
\delta X^7 &=i (\bar{\chi}_R \bar{\epsilon}_L - \chi_R \epsilon_L + \bar{\lambda}_R \bar{\xi}_L - \lambda_R \xi_L)&  \delta_{\epsilon_L} X^7&=  -i\chi_R \epsilon_L +i\bar\chi_R\bar\epsilon_L\\
\delta X^8&=i (\bar{\chi}_L \bar{\epsilon}_L - \chi_L \epsilon_L - \bar{\lambda}_L \bar{\xi}_L + \lambda_L \xi_L)&   \delta_{\epsilon_L} X^8&=  -i\chi_L \epsilon_L + i \bar\chi_L\bar\epsilon_L\\
\delta X^9&=-\bar{\chi}_L \bar{\epsilon}_L - \chi_L \epsilon_L + \bar{\lambda}_L \bar{\xi}_L + \lambda_L \xi_L&  \delta_{\epsilon_L} X^9&=  -\chi_L \epsilon_L  -\bar\chi_L\bar\epsilon_L\\
\delta \Sigma&=i (\bar{\epsilon}_L \bar{\lambda}_R - \epsilon_L \lambda_R - \bar{\chi}_R \bar{\xi}_L + \chi_R \xi_L), &  \delta_{\epsilon_L} \Sigma&=  -i\lambda_R \epsilon_L  +i\bar\lambda_R\bar{\epsilon}_L.
  \end{align}
\end{subequations}
This gives some natural combinations under $\mathcal{N}=2$
\be
\delta_{\epsilon}(X^6+iX^7)=2(\chi_L\epsilon_L-\bar{\lambda}_R\bar{\xi}_L
)\ee
\be
\delta_{\epsilon}(X^8+iX^9)=2i (\lambda_L\xi_L-\chi_L\epsilon_L)
\ee
The gauge field transforms as
\begin{subequations}
  \begin{align}
\mathcal{N}&=2 & \mathcal{N}&=1 \nonumber \\
\delta_{\epsilon_L} \! A_{\mu}&=i (\epsilon_L\sigma^{\mu} \bar{\lambda}_L  \!+\! \bar{\epsilon}_L \bar{\sigma}^{\mu} \lambda_L\! +\! 
 \bar{\xi}_L   \bar{\sigma}^{\mu} \chi_L\!+\!  \xi_L\sigma^{\mu}\bar{\chi}_L  )   &  \delta_{\epsilon_L} A_\mu& = i \epsilon_L\sigma_\mu \bar\lambda_L + i \bar\epsilon_L \bar\sigma_\mu \lambda_L\\
\delta_{\epsilon_{}}A_{4}&=-( \bar{\xi}_{L}\bar{\chi}_{R} +\xi_{L}\chi_{R} )-( \bar{\epsilon}_{L}\bar{\lambda}_{R} +\epsilon_{L}\lambda_{R} ) & \delta_{\epsilon_L} A_4&=-(\epsilon_L\lambda_R +\bar{\epsilon}_L\bar{\lambda}_R) 
  \end{align}
\end{subequations}
The fermions transform under $\mathcal{N}=1$ as
\begin{subequations}
  \begin{align}
    \delta_{\epsilon_L} \lambda_L &=  F_{\mu\nu}\sigma^{\mu\nu}\epsilon_L-iD_4\Sigma\epsilon_L + 2[X^6,X^7]\epsilon_L+2[X^8,X^9]\epsilon_L\\
    \delta_{\epsilon_L} \lambda_R &= -i\sigma^\mu F_{\mu 4} \bar\epsilon_L- D_{\mu}\Sigma \sigma^\mu \bar\epsilon_L + 2 [X^6-iX^7,X^8-iX^9]\epsilon_L\\
    \delta_{\epsilon_L} \chi_L &= -\sigma^\mu D_{\mu}X^8\bar\epsilon_L-i\sigma^\mu D_{\mu}X^9\bar\epsilon_L-2 [X^6-iX^7,\Sigma]\epsilon_L +D_4(X^6-iX^7)]\epsilon_L \\
    \delta_{\epsilon_L} \chi_R &= i\sigma^\mu D_{\mu}X^6\bar\epsilon_L-\sigma^\mu D_{\mu}X^7\bar\epsilon_L-2 i [X^8-iX^9,\Sigma]\epsilon_L +iD_4(X^8-iX^9)]\epsilon_L .
  \end{align}
\end{subequations}
Under $\mathcal{N}=2$ the fermions transform as
 \begin{subequations}
  \begin{align}
\delta_{\epsilon_{}}\lambda_{L}&=+F_{\mu\nu}\sigma^{\mu\nu}\epsilon_{L} -iD_4\Sigma\epsilon_{L} -2[X^6,X^7]\epsilon_L +2[X^8,X^9]\epsilon_L +
\\
   &\xi_L( D_4 X^6- iD_4 X^7)+\sigma^{\mu}\bar{\xi}_LD_{\mu}X^8+i \sigma^{\mu}\bar{\xi}_LD_{\mu}X^9+2 \xi_L [X^6,\Sigma]-2i\xi_L[X^7,\Sigma]\nonumber
\\
\delta_{\epsilon_{}}\lambda_{R}&=
i \sigma^{\mu} F_{4\mu}\bar{\epsilon}_{L} -\sigma^{\mu}D_{\mu}\Sigma \bar{\epsilon}_L  +
D_\mu\Sigma \left(\sigma^\mu {\bar\epsilon}_L\right)-\frac{1}{2}[X^6-iX^7,X^8+iX^9]\epsilon_L 
\\
+&\xi_LD_4 X^8 + \sigma^{\mu}\bar{\xi}_LD_{\mu}X^6-i\sigma^{\mu}\bar{\xi}_lD_{\mu}X^7-
i\xi_LD_5X^9 -2\xi_L[X^8+iX^9,\Sigma] \nonumber
\\
\delta_{\epsilon_{}}\chi_{L}&= D_5(X^6-iX^7)\epsilon_L -\frac{1}{2}[X^6-iX^7,\Sigma]\epsilon_L
 -\frac{i}{2}[\Sigma,X^8-iX^9]\epsilon_L
\\
&\ \ \ \ i\xi_LD_4\Sigma +\xi_L(F_{\mu\nu}\sigma^{\mu\nu}-2[X^6,X^7]+2[X^8,X^9]) \nonumber
\\
\delta_{\epsilon_{}}\chi_{R}&=i D_{\mu}(X^6+iX^7)\sigma^{\mu}\bar{\epsilon}_L - \frac{i}{2}[\Sigma,X^8+iX^9]\epsilon_L 
+ D_5(X^6+iX^7)\epsilon_R
\\
 &-iF_{\mu4}\sigma^{\mu}\bar{\xi}_L+\sigma^{\mu}\bar{\xi}_LD_{\mu}X^6+2\xi_L([X^6,X^8-iX^9]+[X^7,X^8+iX^9]).
\nonumber
 \end{align}
\end{subequations}
This completes our analysis of the orbifolded MSYM theory reduced to $\mathcal{N}=2$ in four dimensional superspace.  Additional conventions may be found in the appendix. The primary purpose of this detailed exposition of the orbifolded MSYM action was that it may have future model building applications. So next we change tone slightly and demonstrate an application of this setup.   
\section{The boundary effective action}\label{Dirac}
In this section we explore applications of maximal super Yang Mills on a  five dimensional orbifold. There are likely to be many uses to the construction of the MSYSM with an orbifold, but we wish to discuss one particular example that makes application also of the boundary terms. In addition all fields $V,H$ and $\Phi, H^c$, with both positive and negative parity play an important role. The application is to supersymmetry breaking in gauge mediation.

In this section we will imagine that the sector that breaks supersymmetry is a strongly coupled system and admits something analogous to an AdS dual.  In addition, for gauge mediated supersymmetry breaking, we expect that this strongly coupled sector  has some subset that is charged under the standard model gauge groups.  
As envisaged in \cite{McGarrie:2010kh,Abel:2010vb,McGarrie:2012fi}, we may wish to imagine a situation in which there is a weakly gauged 
global symmetry $SU(N_F)$ of the strongly coupled system, which will be identified with the standard model gauge groups (or some GUT embedding).  We will model this  system with a bulk slice of $AdS_5$ with extra dimension ranging $L_0<z<L_1$. In those papers  $\mathcal{N}=1$ super Yang-Mills is considered for the bulk fields. Here we  extend this to maximal super Yang-Mills and focus here only on new results not contained in \cite{McGarrie:2012fi}.
The metric is given by
\be
ds^2=a^2(z)(\eta^{\mu\nu}dx_{\mu}dx_{\nu}+dz^2) \ \ \ \text{where}\ \ \   a^2(z) =\left(\frac{R}{z}\right)^2.
\ee
The action of this paper may be extended to warped or AdS space following \cite{Shuster:1999zf,Marti:2001iw,Cacciapaglia:2008bi,Bagger:2011na}.  In particular, consistency with AdS space means that the action contains
\be
\mathcal{L}_{AdS}\supset \left( m_{\psi}\bar{\psi}^I\psi_I + \frac{1}{2}m^2_{X^I}X^IX_I\right).
\ee
These mass terms have 
\be
m_{\psi}R=c  \ \ \  \text{with} \ \ \  \Delta=\frac{3}{2}+|c+\frac{1}{2}|
\ee
$\Delta$ is the scaling dimension and $c$ is a real number which controls the localisation of the field profiles in the $z$ direction ($c=1/2$ is flat).  For the scalars one finds
\be
m^2_{X^I}R=c^2+c-\frac{15}{4} 
\ee
where $c\rightarrow -c$ is also possible \cite{Cacciapaglia:2008bi}, although $\Delta=2$.

The boundary terms of the MSYM theory will now play an important role for us in generating a  boundary effective action. The boundary values of the bulk gauge fields are the sources and we wish to compute the tree level effective action, essentially the correlators of operators that couple to these sources:
\be
\braket{O(p)O(-p)}=\text{Lim}_{L_0\rightarrow 0}(p^2\Pi(p^2)+\text{UV counter terms}).
\ee  
$p,q$ are four dimensional momentum.  For this model, the sources are 
\be
A^0_{\mu} (x),   \lambda^{0}_{\alpha, L}(x) ,D^0(x), \chi^{0}_{\alpha, L}(x), \phi_6^0(x),\phi_7^0(x),\phi_8^0(x),\phi_9^0(x),
\ee
where $x$ is the four dimensional position and the scalar sources are given by
\be
\phi^0_{6,7}= \partial_z X^{6,7} |_{z=L_0}\ \ \, D^0=\partial_zX^{10}|_{z=L_0}  \  \ \text{and}   \ \ \ \phi^0_{8,9}|= X^{8,9}|_{z=L_0}.
\ee
The other fermions $\lambda_R,\chi_{R}$ are free to vary \cite{McGarrie:2012fi}, it is just the sources that are fixed.
 $A^0$, $D^0$, $\lambda^0_L$ and $\chi^0_L$ have even parity.   These source fields are sources for (non) CFT operators. As well as the identification of operators and fields found in \cite{McGarrie:2012fi}, the additional scalars and fermions are identified as in table \ref{table},
\begin{table}
\begin{center}
\begin{tabular}{|c c c | c|c|}
\hline
4D:  operator & & Field& $\Delta$ &$m^2$ \\
\hline
$\mathcal{O}^{I}(x)$&$\rightarrow $&$ X^I(x,z)$ &2  &-4\\
$\mathcal{O}^{\alpha}(x)_{L}$&$\rightarrow $&$  \chi^{\alpha}(x,z)_{L}    $& 5/2 \text{or} 3/2&1/2 \\
\hline
\end{tabular}
\caption{Operators corresponding to the bulk fields of the model.}\label{table}
\end{center}
\end{table}
where in this instance the $L$ on the fermion labels parity under $\gamma^4$ \refe{gamma4} (and should not be confused with a flavour symmetry label).  
 The non vanishing boundary terms at $z=L_0$ are
\be
\frac{1}{g_5^2}\int \frac{d^4p}{(2\pi)^4}\left[ \frac{a(z)}{2}\left(\eta^{\mu\nu}A_{\mu}(p,z)\partial_zA_v(-p,z)-2\eta^{\mu\nu}A_{\mu}(p,z)\partial_{\nu}A_5(-p,z)\right)    \right]+\nonumber
\ee
\be
\frac{1}{g_5^2}\int \frac{d^4p}{(2\pi)^4}\left[ ia^3(z)(\partial_zX_I) X^I +a^4(z)\left(\lambda_L\lambda_R+\chi_L\chi_R+\bar{\lambda}_L\bar{\lambda}_R+\bar{\chi}_L\bar{\chi}_R\right)\right].
\ee
with $X^I, I=6,7,8,9,10$ and $X^{10}=\Sigma$.  One may compute the effective actions for these fields, taking into account the bulk to boundary field profiles and canonical normalisation, 
\be
\lambda^0\rightarrow \lambda^0 a^{3/2}(L_0) \ \ , \ \ \phi^0\rightarrow \phi^0 a(L_0) \ \  , \ \  A^0\rightarrow A^0.
\ee
In particular the bulk fermion field $\lambda_L$ is coupled to $\lambda_R$
\be
\sigma^{\mu}_{\alpha\dot{\alpha}}p_{\mu}\bar{\lambda^0_R}=p\frac{Q_{-}(q,L_0)}{Q_{+}(q,L_0)}\lambda^0_L
\ee
where $Q_{\pm}(q,z)$ are bulk profile functions that are solutions to the bulk fermion equations of motion.  The bulk fields may be decomposed in terms of a source and a bulk profile
\be
(\{\lambda_{L},\chi_L\},\{\lambda_{R},\chi_R\} )(q,z)=\frac{1}{Q_{\pm}(q,L_0)}(\{\lambda^0_{L},\chi^0_L\}_{+},\{\lambda^0_{R},\chi^0_R\}_{-} )Q_{\pm}(q,z),
\ee
suitably normalised by the boundary value of the profile function.   A particular solution is 
\be
Q_+(q,z)=z^{5/2}\left[J_{\alpha}(qz)Y_{\beta}(qL_1)-J_{\beta}(qL_1)Y_{\alpha}(pz)\right],
\ee
with $\alpha=c+1/2$ and $\beta=\alpha-1$. When this is carried out for all the bulk fields one obtains a boundary effective action
\be
\int\!\! \frac{d^4q}{(2\pi)^4}\left[ \Pi_1(q^2)F^{\mu\nu}_0F_{0,\mu\nu}-i\Pi^{\lambda_L}_{1/2}(q^2)\lambda^0_L\sigma^{\mu}\partial_{\mu}\bar{\lambda}^0_L -i\Pi^{\chi_L}_{1/2}(q^2)\chi^0_L\sigma^{\mu}\partial_{\mu}\bar{\chi}^0_L +\frac{1}{2}\Pi_{0}(q^2)D_0^2 \right]\nonumber
\ee
\be
+\int\!\! \frac{d^4q}{(2\pi)^4}\left[ \Pi(q^2)_{\phi_6}(\phi^0_6)^2+\Pi(q^2)_{\phi_7}(\phi^0_7)^2+\Pi(q^2)_{\phi_8}(\phi^0_8)^2+\Pi(q^2)_{\phi_9}(\phi^0_9)^2 \right].
\ee
One should interpret $(\phi^0)^2$ in the above to mean $\phi^0(q)\phi^0(-q)$ etc.
This effective action could not be generated without the additional fields such as $\lambda_R,\chi_{R}$ contained in the MSYSM action.  This is the boundary action for a  four dimensional $\mathcal{N}=2$ SYM $SU(N_F)$ flavour symmetry, in a large $N_c$  strongly coupled system.

The actual form of the correlators turn out to be similar to those found in \cite{McGarrie:2012fi}.
As we have not broken supersymmetry yet, these boundary to boundary correlators of the tree level effective action should cancel, in sets for instance
\be
\left[ \Pi_0(q^2)-4\Pi^{\lambda_{L}}_{1/2}(q^2)+3\Pi_1(q^2)\right]\equiv0.
\ee
But now there is also 
\be
\Pi^{\chi_{L}}_{1/2}(q^2)  \ \ \text{and} \ \  \Pi_0(q^2)_{\phi_I}
\ee
The general form is 
\be
\Pi^{\chi_{L}}_{1/2}(q^2)=\frac{a(z)}{q g_5^2}\left(\frac{Q_{-}(q,z)}{Q_{+}(q,L_0)}\right)_{z=L_0}.
\ee
where the tree level matching condition is given by
\be
\frac{R}{g_5^2}=\frac{N_c}{12\pi^2}.
\ee
It is also interesting to consider that the source-field construction on the UV boundary may be written in superspace,
\be
a^3(L_0)\int d^4 x \int d^2\theta\left[ \Phi^0\Phi(L_0)+ H^c(L_0)H^0+H(L_0)H^{c,0}\right]
\ee
which also incorporate $F^0$ term sources, for instance
\be
 \int d^2\theta\left[  H^cH^0\right]=F_{\mathcal{O}}(\phi_{8}^0+i\phi_{9}^0)+\mathcal{O}^{\alpha}\chi^0_{L,\alpha}+\mathcal{O}F^0_H
\ee
although for our use we find it easier to work with components.

\subsection{Dirac masses from a strongly coupled system}
These results have some interesting applications. Not only does it extend the work of \cite{Contino:2004vy,DaRold:2005ju} in exploring how one may use 5D supersymmetric models to study 4D strongly coupled systems, importantly this setup will allow for Dirac soft masses \cite{Antoniadis:2006uj,Amigo:2008rc,Benakli:2008pg,Benakli:2010gi,Abel:2011dc} to arise in $AdS_5$ between the external source fields $\lambda^0_L$ and $\chi^0_L$ (contained in $V$ and $H_{adj}$).  Extending the current correlator programme of \cite{Benakli:2008pg}, we take an IR localised superpotential 
\be
W=Y_X H_{adj}\mathcal{J}_2
\ee
with $\mathcal{J}_2(y)=J_2+\sqrt{2}\vartheta j_2 +{\vartheta}^2 F_2$, a chiral superfield made of operators that may live in the bulk or IR boundary.
Applying 
\cite{McGarrie:2012fi}, a Dirac soft mass can be interpreted as additional terms in a supersymmetry breaking effective action on the UV boundary, as well as interesting kinetic mixing terms:
\be
S_{eff}\supset\int \frac{d^4p}{(2\pi)^4}\left[g_{SM}\tilde{Y}_X M\tilde{H}_{1/2}(p^2)\lambda^0_L\chi^0_L -ig_{SM}\tilde{Y}_X\tilde{G}_{1/2}(p^2)\chi^0_{L}\sigma_{\mu}\partial^{\mu}\bar{\lambda}^0_L\right]
\ee
\be
+\int \frac{d^4p}{(2\pi)^4}\left[-\frac{\tilde{Y}^2_X}{2}M\tilde{I}_{1/2}(p^2)\chi^0_L\chi^0_L -\frac{i}{2}\tilde{Y}^2_X\tilde{E}_{1/2}(p^2)\chi^0_L\sigma^{\mu}\partial_{\mu}\bar{\chi}^0_L\right]+c.c.
\ee
The fields have been canonically normalised.
We expect that these soft masses are taken in the limit $p^2\rightarrow 0$, such that any dressing functions $\Lambda(p^2)$ (which may incidentally be aborbed into the definition of the Yukawa $\tilde{Y}_X$), associated with the intermediate states, will not suppress this result: $\Lambda(0)=1$ typically.  In summary, this effective action is found by integrating out the bulk states and generate effective Dirac masses between external source fields of fermions $\lambda^0_{L}$and $\chi^0_{L}$ .
\subsection{Cross sections for Dirac soft masses}
In \cite{Fortin:2011ad} cross sections of visible to hidden sector matter were considered for a straightforward messenger model.  These techniques may be applied to the correlators associated with Dirac soft masses, and moreover, if the hidden sector is strongly coupled these correlators may develop certain form factors as documented in \cite{McGarrie:2012ks,McGarrie:2012fi}.  For these reasons we explore the cross-sections for the Dirac soft mass correlators.   In general these cross sections will be valid for both $\sigma(\lambda_L\rightarrow \text{hidden})$ or $\sigma(\chi_L\rightarrow \text{hidden}) $.

If a hidden U(1) gauge field develops a vev $W'_{\alpha}=\theta_{\alpha}D$  the messenger fields $Q,\tilde{Q}$ with opposite charges under the U(1) are split $M^2\pm D$ with a current \cite{Benakli:2008pg}
\be
\mathcal{J}_2=Q\tilde{Q}
\ee
coupled to an adjoint chiral superfield
\be
W\supset Y_{X}H_{adj}\mathcal{J}_2
\ee
then one obtains a Dirac soft mass between $\lambda_L$ and $\chi_L$
\be
m_D=g Y_X M\tilde{H}_{1/2}(0).
\ee
As the function 
$\tilde{H}_{1/2}(p^2/M^2)$ is structurally the same as $M\tilde{B}_{1/2}$  we may compute the cross section  to be
\be
\sigma_{D}=\frac{(4\pi)\alpha_{X}}{m_0 s}\text{Im}\left[iM\tilde{H}_{1/2}(s)\right]=\frac{(4\pi)\alpha_{X}}{m_0 s}\frac{1}{2i}\text{Disc}\left[iM\tilde{H}_{1/2}(s)\right]
\ee
where we have defined $\alpha_{X}=g_{SM}Y_X/4\pi$ and
\be
\text{Disc}\left[iM\tilde{H}_{1/2}(s)\right]=\frac{m_0}{4\pi s}\lambda^{1/2}(s,m^2_0,m^2_+)\theta(s-(m_0+m_+)^2) - (m_+\rightarrow m_-)
\ee
In the result above we have used some notation. We have introduce the `triangle function':
\be 
\lambda(s,m_1^2 ,m^2_2)=4 s |\bold{p}|^2= (s^2+m^4_1 +m^4_2)- 2 s m_1^2 - 2s m_2^2- 2m^2_1 m^2_2
\ee
where  $s$ is the centre of mass energy squared Mandelstam variable $s=(p_1+p_2)^2$.  
Multiplying by 
\be
\theta(s-(m_1+m_2)^2)
\ee
which is equivalent to 
\be
\lambda^{1/2}(s,m^2_1,m^2_2)=2\sqrt{s}|p^{ \! \! \! \to}|=\sqrt{[s-(m_1+m_2)^2][s-(m_1-m_2)^2]}\theta(s-(m_1+m_2)^2).
\ee 
For equal mass scattering it has a very simple form
\be
\lambda^{1/2}(s,m,m)=[s(s-4m^2)]^{1/2}=s\sqrt{1-\frac{4m^2}{s}}
\ee
and for unequal masses with  $m_0=M, \ \  m^2_{\pm}=M^2\pm F$ (or similarly, $m^2_{\pm}=M^2\pm D$) one gets
\be
\lambda^{1/2}(s,m_0,m_+)=\sqrt{s^2+F^2-2s (m_0^2+m_+^2)}
\ee
\be
\lambda^{1/2}(s,m_0,m_+)=\sqrt{s^2+F^2-2s (m_0^2+m_-^2)}
\ee
\be
\lambda^{1/2}(s,m_-,m_+)=\sqrt{s^2-4m_0^2 s+4F^2}.
\ee
Conversely, for a model with an R-symmetric F term \cite{Benakli:2008pg}, the Dirac mass is given by 
\be
\tilde{H}_{1/2}(p^2/M^2)=\frac{1}{\sqrt{2}}M \cos (\xi/v)\int \frac{d^4 q}{(2\pi)^4}\frac{1}{(q+p)^2+M^2}
\ee
\be
\times  \left(\frac{1}{q^2+m^2_+}+\frac{1}{q^2+m_-^2}-\frac{2}{q^2+M^2} \right).     \nonumber
\ee
This will give a cross section 
\be
\text{Disc}\left[iM\tilde{H}_{1/2}(s)\right]=
\ee
\be
\frac{\cos(\xi/v)}{\sqrt{2}}\frac{M}{4\pi s}\left[ \lambda^{1/2}(s,m_0^2,m_+^2)\theta+  \lambda^{1/2}(s,m_0^2,m_-^2)\theta - 2\lambda^{1/2}(s,m^2_0,m^2_0) \theta  \right]. \nonumber
\ee
These cross sections may be dressed by the appropriate form factor squared $|F(s)|^2$
\be
F(s)=\frac{Q_{-}(s,L_1)}{Q_{+}(s,L_0)}=-g_5\sum_{n=1} \frac{F_n\psi_n(L_1)}{s-m_n^2}
\ee
associated with the bulk to boundary propagator between the UV boundary and the IR brane.
We hope that these ideas have aided in extending the Dirac soft mass programme \cite{Benakli:2008pg,Abel:2011dc} also to certain strongly coupled systems. 
\section{Discussion and conclusions}\label{Conclude}
In this paper we have determined the superspace formulation of maximal super Yang-Mills in five dimensions, on a manifold with boundaries.  This type of setup perhaps offers a new resolution of the issue of non renormalisability of five dimensional model building  \cite{Douglas:2010iu,Lambert:2010iw}.     In addition it allows for a positive parity vector $V$ and chiral field $H_{adj}$, consistent with supersymmetry and dimensional reduction, which may allow for Dirac soft masses in a natural manner.

 In addition we have looked at the boundary terms that result from closure of supersymmetry of the action.   This motivated us to discuss various examples where the boundary action, and necessity of descending from the full MSYM, allows one to compute the complete boundary to boundary correlators and Dirac soft masses between external sources, in a slice of $AdS_5$.

In particular, by applying the holographic interpretation of this setup, we have shown that some fermionic source fields $\lambda^0_{L}$ and $\chi^0_{L}$ which couple to some (non) CFT operators $\mathcal{O}_{L,R}$, may develop Dirac soft masses after the bulk action is completely integrated out.  This setup can be straightforwardly extended to more general or more complicated AdS models. It is natural to also write this action in a fully warped superspace  following \cite{Marti:2001iw,Cacciapaglia:2008bi,Bagger:2011na}.   It could also be possible to construct an entirely four dimensional quiver Deconstruction of this setup,  which is likely to be more palatable to some. Finally it might be interesting to see M-theory play a more prominent role in phenomenology.

\paragraph{Acknowledgements} 
I am funded by the Alexander von Humboldt Foundation. I am especially greatful for the assistance of \"Omer G\"urdo\u gan, as well as Daniel C. Thompson, Ingo Kirsch, Steven Abel  and Neil Lambert for useful discussions or comments. I would also like to acknowledge the warm hospitality of the IPPP-Durham, whilst some of this work was completed.
\appendix

\section{Conventions and spinors}\label{conventions}
In this appendix we outline the conventions used.  The indices $\mu,\nu$ are 4D , $\mu=0,1,2,3$.  $M,N$ are 5D indices with metric $\eta_{MN}=\text{diag}(-1,1,1,1,1)$.  $A,B$ are the 11D indices.  $I,J=6,7,8,9,10$.  In particular the $\Gamma^{\hat{M}\hat{N}}$ multiply 7d spinors, the $\gamma^{MN}$ multiply symplectic Majorana spinors and the $\sigma^{MN}$ multiply  2-component spinors.\\

We define the 11d Gamma matrices to satisfy the Clifford algebra
\be
\{\Gamma^A, \Gamma^B\}=-2\eta^{AB}\mathbb{1}
\ee
\be
(\Gamma^A)^T=C_{11}\Gamma^A C_{11}^{-1}
\ee
\be
(\Gamma^A)^*=B_{11}\Gamma^A B_{11}^{-1}
\ee
\be
B_{11}=C_{11}\Gamma^0.
\ee
To dimensionally reduce to four dimensions we choose the explicit representation 
\be
\Gamma^{M}= \sigma_3 \otimes \sigma_3 \otimes \sigma_3 \otimes \gamma^{M}
\ee
\be
\Gamma^{6}= \sigma_3 \otimes  \sigma_3 \otimes  \sigma_1 \otimes \mathbb{1}_4
\ee
\be
\Gamma^{7}= \sigma_3 \otimes  \sigma_3 \otimes  \sigma_2 \otimes \mathbb{1}_4
\ee
\be
\Gamma^{8}= \sigma_2\otimes  \mathbb{1}_2 \otimes  \mathbb{1}_2 \otimes \mathbb{1}_4
\ee
\be
\Gamma^{9}=- \sigma_1 \otimes \mathbb{1}_2 \otimes  \mathbb{1}_2 \otimes \mathbb{1}_4
\ee
\be
\Gamma^{10}= \sigma_3 \otimes  \sigma_1 \otimes \mathbb{1}_2  \otimes \mathbb{1}_4
\ee
\be
\Gamma^{5}= \sigma_3 \otimes  \sigma_2 \otimes \mathbb{1}_2 \otimes \mathbb{1}_4.
\ee
$\Gamma^5$ should satisfy:
\be
\Gamma^{5}= i \Gamma^0 \Gamma^1...\Gamma^9\Gamma^{10}.
\ee
In two component spinor notation
\be
\gamma^M=\left(\,\left(\begin{array}{cc}0&\sigma^\mu_{\alpha \dot{\alpha}}\\ 
\bar{\sigma}^{\mu \dot{\alpha} \alpha }&0
\end{array}\right),
\left(\begin{array}{cc}-i&0\\ 0&i\end{array}\right)\,
\right)\,,~~\mbox{and}~~~
C_5=\left(\begin{array}{cc}
-\epsilon_{\alpha\beta} & 0\\ 
0 & \epsilon^{\dot\alpha \dot\beta}
\end{array}\right)\,,
\ee
where $\sigma^\mu_{\alpha \dot{\alpha}}=(1,\vec{\sigma})$ and
$\bar{\sigma}^{\mu \dot{\alpha}
\alpha}=(1,-\vec{\sigma})$. $\alpha,\dot{\alpha}$ are spinor indices
of $\text{SL}(2,C)$.   The $\gamma^4_{5d}=-i \gamma^5_{4d}$  where explicitly
\be
\gamma^5_{4d}=\left(\begin{array}{cc}
\mathbf{I}& 0\\ 
0 &-\mathbf{I}
\end{array}\right)\,.\label{gamma4}
\ee
This may also be written as $\gamma^5_{4d}=i\gamma^0\gamma^1\gamma^2\gamma^4 \gamma^5$.  $C_5$ is the 5d charge conjugation matrix such that 
\be
C_5\gamma^M C^{-1}_5=\left(\gamma^M\right)^T.
\ee
We may also define the complex conjugation matrix 
\be
B_5=C_5\gamma^0=\left(\begin{array}{cc}0&
-\epsilon^{\alpha\beta} \\ 
\epsilon_{\dot\alpha \dot\beta}&0
\end{array}\right)\,.
\ee
We include also the Pauli matrices 
\be
\sigma_1=\left(\begin{array}{cc}
0& 1\\ 
1 &0
\end{array}\right) \ \ 
\sigma_2=\left(\begin{array}{cc}
0& -i\\ 
i &0
\end{array}\right) \ \ 
\sigma_3=\left(\begin{array}{cc}
1& 0\\ 
0 &-1
\end{array}\right),
\ee
such that one may verify the tensor products.
\subsection{Seven dimensions}
For obtaining the 5d Lagrangian from seven-dimensional $\mathcal{N}=1$ supersymmetric Yang-Mills theory \cite{Ludeling:2011ip}, the following set of gamma matrices is particularly convenient:
\begin{equation}
  \Gamma^{\hat M}=\left\{\sigma^3\otimes\gamma^M,\,\sigma^1\otimes\mathbb{1}_4,\,\sigma^2\otimes\mathbb{1}_4\right\}.
\end{equation}
Note that the five dimensional matrices are embedded in the first five components of $\Gamma^{\hat M}$. Moreover, $\Gamma^5$ and $\Gamma^6$ act as the identity operator to the two halves of the eight-component spinors of seven dimensions.

Symplectic Majorana spinors in seven dimensions are defined as follows:
\begin{equation}
  \label{eq:7dSM}
  \psi_I=\epsilon_{IJ}C_{7}\left({\bar\psi}^T\right)^J, \ \  \psi_I=\epsilon_{IJ}B_{7}\left({\psi^*}\right)^J,  
\end{equation}
with the seven-dimensional charge conjugation matrix being
\begin{equation}
  C_{7}=i\Gamma^{\hat{0}}\Gamma^{\hat{2}}\Gamma^{\hat{4}}\Gamma^{\hat{5}}=\sigma^2\otimes\mathbb{1}_2\otimes\sigma^2
=-i\sigma^2\otimes C_5
\end{equation}
and  complex conjugation is defined as
\begin{equation}
  B_{7}=C_7 \Gamma^{\hat{0}} .
\end{equation}
In this basis, a pair of seven-dimensional symplectic Majorana spinors satisfying \eqref{eq:7dSM} have the form\begin{equation}
\psi_1=\left(
  \begin{array}{c}
    \Omega^1\\
    \Psi^1
  \end{array}
\right),\qquad
\psi_2=\left(
  \begin{array}{c}
    \Psi^2\\
    \Omega^2
  \end{array}
\right), \qquad
\epsilon_1=\left(
  \begin{array}{c}
    \xi^1\\
    \epsilon^1
  \end{array}
\right),\qquad
\epsilon_2=\left(
  \begin{array}{c}
    \epsilon^2\\
    \xi^2
  \end{array}
\right),
\end{equation}
 where
$(\Omega^1,\Omega^2)$ and $(\Psi^1, \Psi^2)$ are pairs of four component spinors, separately satisfying the five-dimensional symplectic Majorana condition:
\begin{equation}
  \Omega^i=\epsilon^{ij}C{\bar\Omega}^T_j,\qquad  \Psi^i=\epsilon^{ij}C{\bar\Psi}^T_j.
\end{equation}
The 7d supersymmetry parameters may be written
\be 
(\bar{\eta}^1)^T=\left(
  \begin{array}{c}
    \bar{\xi}_1\\
    -\bar{\epsilon}_1
  \end{array}
\right),\qquad
(\bar{\eta}^2)^T=\left(
  \begin{array}{c}
    \bar{\epsilon}_2\\
    -\bar{\xi}_2
  \end{array}
\right).
\ee
It is useful to identify  the degrees of freedom of the 7d and 5d theories
\be
(B_1+iB_2)= (X_8+iX_9 )  \ \  \text{and}  \ \ B_3= X_{10}= \Sigma,
\ee
where $\Sigma$ is a real adjoint scalar matching the notation of $\mathcal{N}=1$ 5d SYM.  The seven dimensional SYM boundary action ( of $x_4$) is given by 
\be
\frac{1}{g_5^2}\int  d^4 x \left( G^{4M} G^{PQ}F_{MP}A_{Q}-\frac{1}{2}G^{4M}(D_M B_i)B^i +\frac{1}{4}\bar{\psi}^I \psi_I \right)
\ee
which are naturally contained in the superfield, but not the component action of \cite{Ludeling:2011ip}.


\bibliographystyle{JHEP}
\bibliography{MAXI}

\providecommand{\href}[2]{#2}\begingroup\raggedright\begin{thebibliography}{10}

\bibitem{Douglas:2010iu}
M.~R. Douglas, {\it {On D=5 super Yang-Mills theory and (2,0) theory}},  {\em
  JHEP} {\bf 1102} (2011) 011, [\href{http://xxx.lanl.gov/abs/1012.2880}{{\tt
  arXiv:1012.2880}}].

\bibitem{Lambert:2010iw}
N.~Lambert, C.~Papageorgakis, and M.~Schmidt-Sommerfeld, {\it {M5-Branes,
  D4-Branes and Quantum 5D super-Yang-Mills}},  {\em JHEP} {\bf 01} (2011) 083,
  [\href{http://xxx.lanl.gov/abs/1012.2882}{{\tt arXiv:1012.2882}}].

\bibitem{Horava:1996ma}
P.~Horava and E.~Witten, {\it {Eleven-dimensional supergravity on a manifold
  with boundary}},  {\em Nucl.Phys.} {\bf B475} (1996) 94--114,
  [\href{http://xxx.lanl.gov/abs/hep-th/9603142}{{\tt hep-th/9603142}}].

\bibitem{Mirabelli:1997aj}
E.~A. Mirabelli and M.~E. Peskin, {\it {Transmission of supersymmetry breaking
  from a four-dimensional boundary}},  {\em Phys.Rev.} {\bf D58} (1998) 065002,
  [\href{http://xxx.lanl.gov/abs/hep-th/9712214}{{\tt hep-th/9712214}}].

\bibitem{Randall:1999ee}
L.~Randall and R.~Sundrum, {\it {A Large mass hierarchy from a small extra
  dimension}},  {\em Phys.Rev.Lett.} {\bf 83} (1999) 3370--3373,
  [\href{http://xxx.lanl.gov/abs/hep-ph/9905221}{{\tt hep-ph/9905221}}].

\bibitem{Erlich:2005qh}
J.~Erlich, E.~Katz, D.~T. Son, and M.~A. Stephanov, {\it {QCD and a holographic
  model of hadrons}},  {\em Phys.Rev.Lett.} {\bf 95} (2005) 261602,
  [\href{http://xxx.lanl.gov/abs/hep-ph/0501128}{{\tt hep-ph/0501128}}].

\bibitem{DaRold:2005ju}
L.~Da~Rold and A.~Pomarol, {\it {Chiral symmetry breaking from five dimensional
  spaces}},  {\em PoS} {\bf HEP2005} (2006) 355.

\bibitem{Abel:2011dc}
S.~Abel and M.~Goodsell, {\it {Easy Dirac Gauginos}},  {\em JHEP} {\bf 1106}
  (2011) 064, [\href{http://xxx.lanl.gov/abs/1102.0014}{{\tt
  arXiv:1102.0014}}].

\bibitem{Benakli:2012cy}
K.~Benakli, M.~D. Goodsell, and F.~Staub, {\it {Dirac Gauginos and the 125 GeV
  Higgs}},  \href{http://xxx.lanl.gov/abs/1211.0552}{{\tt arXiv:1211.0552}}.

\bibitem{Bhattacharyya:2012qj}
G.~Bhattacharyya and T.~S. Ray, {\it {Pushing the SUSY Higgs mass towards 125
  GeV with a color adjoint}},  {\em Phys.Rev.} {\bf D87} (2013) 015017,
  [\href{http://xxx.lanl.gov/abs/1210.0594}{{\tt arXiv:1210.0594}}].

\bibitem{Heikinheimo:2011fk}
M.~Heikinheimo, M.~Kellerstein, and V.~Sanz, {\it {How Many Supersymmetries?}},
   {\em JHEP} {\bf 1204} (2012) 043,
  [\href{http://xxx.lanl.gov/abs/1111.4322}{{\tt arXiv:1111.4322}}].

\bibitem{Hebecker:2001ke}
A.~Hebecker, {\it {5D super Yang-Mills theory in 4-D superspace, superfield
  brane operators, and applications to orbifold GUTs}},  {\em Nucl. Phys.} {\bf
  B632} (2002) 101--113, [\href{http://xxx.lanl.gov/abs/hep-ph/0112230}{{\tt
  hep-ph/0112230}}].

\bibitem{ArkaniHamed:2001tb}
N.~Arkani-Hamed, T.~Gregoire, and J.~G. Wacker, {\it {Higher dimensional
  supersymmetry in 4-D superspace}},  {\em JHEP} {\bf 0203} (2002) 055,
  [\href{http://xxx.lanl.gov/abs/hep-th/0101233}{{\tt hep-th/0101233}}].

\bibitem{Ludeling:2011ip}
C.~Ludeling, {\it {Seven-Dimensional Super-Yang-Mills Theory in N=1
  Superfields}},  \href{http://xxx.lanl.gov/abs/1102.0285}{{\tt
  arXiv:1102.0285}}.

\bibitem{McGarrie:2010yk}
M.~McGarrie and D.~C. Thompson, {\it {Warped General Gauge Mediation}},  {\em
  Phys.Rev.} {\bf D82} (2010) 125034,
  [\href{http://xxx.lanl.gov/abs/1009.4696}{{\tt arXiv:1009.4696}}].

\bibitem{McGarrie:2010kh}
M.~McGarrie and R.~Russo, {\it {General Gauge Mediation in 5D}},  {\em
  Phys.Rev.} {\bf D82} (2010) 035001,
  [\href{http://xxx.lanl.gov/abs/1004.3305}{{\tt arXiv:1004.3305}}].

\bibitem{Abel:2010vb}
S.~Abel and T.~Gherghetta, {\it {A slice of AdS5 as the large N limit of
  Seiberg duality}},  {\em JHEP} {\bf 1012} (2010) 091,
  [\href{http://xxx.lanl.gov/abs/1010.5655}{{\tt arXiv:1010.5655}}].

\bibitem{McGarrie:2011av}
M.~McGarrie, {\it {Gauge Mediated Supersymmetry Breaking in Five Dimensions}},
  \href{http://xxx.lanl.gov/abs/1109.6245}{{\tt arXiv:1109.6245}}. Ph.D. Thesis
  (Advisors: Steven Thomas and Rodolfo Russo).

\bibitem{McGarrie:2012fi}
M.~McGarrie, {\it {Holography for General Gauge Mediation}},  {\em JHEP} {\bf
  1302} (2013) 132, [\href{http://xxx.lanl.gov/abs/1210.4935}{{\tt
  arXiv:1210.4935}}].

\bibitem{ArkaniHamed:2001ie}
N.~Arkani-Hamed, A.~G. Cohen, D.~B. Kaplan, A.~Karch, and L.~Motl, {\it
  {Deconstructing (2,0) and little string theories}},  {\em JHEP} {\bf 0301}
  (2003) 083, [\href{http://xxx.lanl.gov/abs/hep-th/0110146}{{\tt
  hep-th/0110146}}].

\bibitem{McGarrie:2010qr}
M.~McGarrie, {\it {General Gauge Mediation and Deconstruction}},  {\em JHEP}
  {\bf 1011} (2010) 152, [\href{http://xxx.lanl.gov/abs/1009.0012}{{\tt
  arXiv:1009.0012}}].

\bibitem{McGarrie:2011dc}
M.~McGarrie, {\it {Hybrid Gauge Mediation}},  {\em JHEP} {\bf 1109} (2011) 138,
  [\href{http://xxx.lanl.gov/abs/1101.5158}{{\tt arXiv:1101.5158}}].

\bibitem{Lambert:2012qy}
N.~Lambert, C.~Papageorgakis, and M.~Schmidt-Sommerfeld, {\it {Deconstructing
  (2,0) Proposals}},  \href{http://xxx.lanl.gov/abs/1212.3337}{{\tt
  arXiv:1212.3337}}.

\bibitem{Belyaev:2005rs}
D.~V. Belyaev, {\it {Boundary conditions in the Mirabelli and Peskin model}},
  {\em JHEP} {\bf 0601} (2006) 046,
  [\href{http://xxx.lanl.gov/abs/hep-th/0509171}{{\tt hep-th/0509171}}].

\bibitem{Belyaev:2006jg}
D.~V. Belyaev, {\it {Supersymmetric bulk-brane coupling with odd gauge
  fields}},  {\em JHEP} {\bf 0608} (2006) 032,
  [\href{http://xxx.lanl.gov/abs/hep-th/0605282}{{\tt hep-th/0605282}}].

\bibitem{Berman:2009kj}
D.~S. Berman and D.~C. Thompson, {\it {Membranes with a boundary}},  {\em
  Nucl.Phys.} {\bf B820} (2009) 503--533,
  [\href{http://xxx.lanl.gov/abs/0904.0241}{{\tt arXiv:0904.0241}}].

\bibitem{Berman:2011kg}
D.~S. Berman, E.~T. Musaev, and M.~J. Perry, {\it {Boundary Terms in
  Generalized Geometry and doubled field theory}},  {\em Phys.Lett.} {\bf B706}
  (2011) 228--231, [\href{http://xxx.lanl.gov/abs/1110.3097}{{\tt
  arXiv:1110.3097}}].

\bibitem{Arutyunov:1998ve}
G.~Arutyunov and S.~Frolov, {\it {On the origin of supergravity boundary terms
  in the AdS / CFT correspondence}},  {\em Nucl.Phys.} {\bf B544} (1999)
  576--589, [\href{http://xxx.lanl.gov/abs/hep-th/9806216}{{\tt
  hep-th/9806216}}].

\bibitem{Kirsch:2003kx}
I.~Kirsch and D.~Oprisa, {\it {Towards the deconstruction of M theory}},  {\em
  JHEP} {\bf 0401} (2004) 020,
  [\href{http://xxx.lanl.gov/abs/hep-th/0307180}{{\tt hep-th/0307180}}].

\bibitem{Kirsch:2004km}
I.~Kirsch, {\it {Generalizations of the AdS / CFT correspondence}},  {\em
  Fortsch.Phys.} {\bf 52} (2004) 727--826,
  [\href{http://xxx.lanl.gov/abs/hep-th/0406274}{{\tt hep-th/0406274}}].

\bibitem{Shuster:1999zf}
E.~Shuster, {\it {Killing spinors and supersymmetry on AdS}},  {\em Nucl.Phys.}
  {\bf B554} (1999) 198--214,
  [\href{http://xxx.lanl.gov/abs/hep-th/9902129}{{\tt hep-th/9902129}}].

\bibitem{Marti:2001iw}
D.~Marti and A.~Pomarol, {\it {Supersymmetric theories with compact extra
  dimensions in N=1 superfields}},  {\em Phys.Rev.} {\bf D64} (2001) 105025,
  [\href{http://xxx.lanl.gov/abs/hep-th/0106256}{{\tt hep-th/0106256}}].

\bibitem{Cacciapaglia:2008bi}
G.~Cacciapaglia, G.~Marandella, and J.~Terning, {\it {Dimensions of
  Supersymmetric Operators from AdS/CFT}},  {\em JHEP} {\bf 0906} (2009) 027,
  [\href{http://xxx.lanl.gov/abs/0802.2946}{{\tt arXiv:0802.2946}}].

\bibitem{Bagger:2011na}
J.~Bagger and C.~Xiong, {\it {$AdS_5$ Supersymmetry in N=1 Superspace}},  {\em
  JHEP} {\bf 1107} (2011) 119, [\href{http://xxx.lanl.gov/abs/1105.4852}{{\tt
  arXiv:1105.4852}}].

\bibitem{Contino:2004vy}
R.~Contino and A.~Pomarol, {\it {Holography for fermions}},  {\em JHEP} {\bf
  0411} (2004) 058, [\href{http://xxx.lanl.gov/abs/hep-th/0406257}{{\tt
  hep-th/0406257}}].

\bibitem{Antoniadis:2006uj}
I.~Antoniadis, K.~Benakli, A.~Delgado, and M.~Quiros, {\it {A New gauge
  mediation theory}},  {\em Adv.Stud.Theor.Phys.} {\bf 2} (2008) 645--672,
  [\href{http://xxx.lanl.gov/abs/hep-ph/0610265}{{\tt hep-ph/0610265}}].

\bibitem{Amigo:2008rc}
S.~D.~L. Amigo, A.~E. Blechman, P.~J. Fox, and E.~Poppitz, {\it {R-symmetric
  gauge mediation}},  {\em JHEP} {\bf 0901} (2009) 018,
  [\href{http://xxx.lanl.gov/abs/0809.1112}{{\tt arXiv:0809.1112}}].

\bibitem{Benakli:2008pg}
K.~Benakli and M.~Goodsell, {\it {Dirac Gauginos in General Gauge Mediation}},
  {\em Nucl.Phys.} {\bf B816} (2009) 185--203,
  [\href{http://xxx.lanl.gov/abs/0811.4409}{{\tt arXiv:0811.4409}}].

\bibitem{Benakli:2010gi}
K.~Benakli and M.~Goodsell, {\it {Dirac Gauginos, Gauge Mediation and
  Unification}},  {\em Nucl.Phys.} {\bf B840} (2010) 1--28,
  [\href{http://xxx.lanl.gov/abs/1003.4957}{{\tt arXiv:1003.4957}}].

\bibitem{Fortin:2011ad}
J.-F. Fortin, K.~Intriligator, and A.~Stergiou, {\it {Superconformally
  Covariant OPE and General Gauge Mediation}},  {\em JHEP} {\bf 1112} (2011)
  064, [\href{http://xxx.lanl.gov/abs/1109.4940}{{\tt arXiv:1109.4940}}].

\bibitem{McGarrie:2012ks}
M.~McGarrie, {\it {General Resonance Mediation}},  {\em JHEP} {\bf 1303} (2013)
  093, [\href{http://xxx.lanl.gov/abs/1207.4484}{{\tt arXiv:1207.4484}}].

\end{thebibliography}\endgroup

\end{document}